\documentstyle[12pt,epsfig]{article}
\begin{document}
\title{ \vspace{-12mm}
       {\normalsize
       \begin{tabbing}
       \'August 1993   \`hep-th/9308131 \\
       \'DESY 93--118
       \end{tabbing} }
       \vspace{8mm}
Cosmological String Solutions in 4 Dimensions from 5d Black Holes}

\renewcommand{\thefootnote}{\fnsymbol{footnote}}

\author{
K. Behrndt\thanks{e-mail: behrndt@r6rz.ifh.de} \\
{\normalsize \em DESY-Institut f\"ur Hochenergiephysik, Zeuthen}\\
\setcounter{footnote}{6}
S. F\"orste\thanks{e-mail: foerste@convex.ifh.de, supported by DFG,
address after August:
Racah Institute of Physics, The Hebrew University of Jerusalem
}\\
{\normalsize \em Humboldt-Universit\"at, Berlin}}
\date{}
\maketitle
\begin{abstract}
We obtain cosmological four dimensional solutions of the low energy effective
string theory by reducing a five dimensional black hole, and
black hole--de Sitter solution of Einstein gravity down to four dimensions.
The appearance of a cosmological constant in the five dimensional
Einstein--Hilbert action produces a special dilaton potential in the four
dimensional effective string action. Cosmological scenarios implemented by
our solutions are discussed.
\end{abstract}

\renewcommand{\arraystretch}{1.6}
\renewcommand{\thefootnote}{\alph{footnote}}
\newcommand{\pa}{\partial}
\newcommand{\be}{\begin{equation}}
\newcommand{\ee}{\end{equation}}
\newcommand{\ba}{\begin{array}}
\newcommand{\ea}{\end{array}}
\newcommand{\aaa}{\alpha}
\newcommand{\eee}{\epsilon}
\newcommand{\lll}{\lambda}
\newcommand{\LL}{\Lambda}
\newcommand{\aap}{\alpha^{\prime}}
\newcommand{\sss}{\mbox{$\sigma$}}
\newcommand{\bbb}{\beta}
\newcommand{\vsf}{\vspace{5mm}}

To describe strings propagating in nontrivial background
there are two different approaches. First, one can start
with a conformal field theory, e.g.~(gauged) WZW, Feigin-Fuchs or Liouville
theory, and rewrite this as a string $\sss$ model.
Although the gauged WZW model is an exact conformal field theory
it is possible to obtain the $\sigma $ model background only perturbatively.
The reason is that in general
the 2d gauge field can be eliminated only in an $\aap$ expansion.
One example for which
one got an exact result is the $SL(2,{\bf R})/U(1)$ WZW model
\cite{witt,verl}. In the second approach one starts with the low energy
effective action and tries to find solutions in an
$\aap$ expansion. In terms of both approaches some
cosmological solutions have been obtained so far. Exact solutions, e.g.~,
are given by the gauged WZW model
based on $SL(2,{\bf R})/SO(1,1) \times {\bf R}^2$ \cite{luest} or
the combination of the $SU(2)$ WZW model with a Feigin-Fuchs part
\cite{ellis,behr}. In the second or ``phenomenological'' approach
cosmological solutions have
been discussed in \cite{tseyt} and \cite{muell}. 
In all these solutions the metric is either time independent like 
the $SU(2)$ WZW in the string frame\footnote{This corresponds to a 
linear expanding universe in the Einstein frame (see \cite{ellis})} 
or spatially flat or the dilaton is constant. In the present
paper we use the second approach and find time dependent solutions 
(metric and dilaton) which are spatially not flat. We use the
following procedure. We neglect all terms ${\cal O}(\alpha'^2)$ and
consider only curvature and dilaton terms. Then we will rewrite the
effective string action in four dimensions (4d) as a 5d
Einstein-Hilbert action and look for solutions of this 5d
theory. After reducing back to 4d we obtain a solution of the 
conformal invariance conditions ($\bar{\bbb }$ equations) of the
corresponding $\sigma$ model.

\vsf

So, restricting in the following on curvature and dilaton terms
the 4d effective action in the lowest order in $\alpha'$ is given by
\be                                             \label{action1}
S=\frac{1}{2}\int d^4 x \sqrt{|G|}\, e^{-2\phi}\left(R +
                    4 (\pa \phi)^2 \right) ~,
\ee
where we assume that the central charge term is zero, i.e.\ there is also 
a decoupled conformal field theory describing the compactified dimensions.
The action (\ref{action1}) is a special case ($\omega = -1$) of the more
general Jordan-Brans-Dicke (JBD) theory,
\be                        \label{jbd}
\begin{array}{rcl}
S&=&\frac{1}{2}\int \sqrt{|G|} \left( \eta R -
    \omega \eta^{-1}(\pa \eta)^2\right)\\
 &=&\frac{1}{2} \int \sqrt{|G|}\,e^{-2\phi} \left( R -
        4\omega (\pa \phi)^2\right)
\ea
\ee
for $\eta = e^{-2\phi}$. In this theory there are some observational
restrictions coming from radar time delay and nucleosynthesis
\cite{scherk,tseyt2}: $|\omega|$ should be larger than a few hundreds
which is certainly not the case for models motivated by string
$\sigma $ models ($\omega = -1$).
A solution of this shortcoming could be provided by the incorporation
of non-perturbative contributions like nontrivial dilaton potentials.
However, the analogy between the string effective action and  JBD theory is
valid only at this level. If we couple further matter fields or
if we consider higher order terms there is a crucial difference.
In string theory matter fields couple directly to the dilaton field
(via $e^{-2\phi}$), whereas in the
JBD theory further matter fields are decoupled  from the scalar field $\eta$.

Let us now transform this 4d JBD action into a 5d
Einstein-Hilbert (EH) action. First we define a five dimensional
metric (not depending on the fifth coordinate),
\renewcommand{\arraystretch}{0.8}
\be                                  \label{fuenf}
G_{\mu\nu}  \rightarrow \tilde{G}_{ab} =
  \left(
     \begin{array}{c|c} e^{(\alpha + \beta)\phi} &\\
                 \hline & \\
                         & \  e^{\beta \phi}\, G_{\mu\nu} \  \\ &
        \end{array} \right),
\ee
\renewcommand{\arraystretch}{1.6}
with
\begin{equation}
\alpha = 2 \left(1 \pm 3\sqrt{1 + \frac{2}{3} \omega}\right), \qquad
\beta = -2 \left(1 \pm \sqrt{1 + \frac{2}{3} \omega}\right), \label{alf}
\end{equation}
and Latin (Greek) indices are running from one to five (four). Using the
five dimensional metric (\ref{fuenf}) and adding one dummy integration
we get for the JBD action (\ref{jbd}) the  5d EH action
\be
S \rightarrow \tilde{S} = \int d^5 x \sqrt{|\tilde{G}|} \, \tilde{R} ~.
\ee
A similar procedure for constructing charged BH solutions is described
in \cite{horo}.
Before we use this procedure for the construction of new solutions
we want to discuss some general features of cosmological solutions.
In cosmology one starts with the spatially isotropic and homogeneous
ansatz (Robertson-Walker),
\be
\begin{array}{l}
ds^2 = -d\tau^2 + K^2(\tau)\left[ dr^2 + f^2(r) \left(
                    \sin^2\theta d\phi^2 + d\theta^2\right) \right]
  = -d\tau^2 + K^2(\tau) d\Omega_{3,\epsilon}^{2} ,\\
          f(r) = \left\{ \begin{array}{rlrr}
             \sin r \qquad & \mbox{for} \qquad & \epsilon = +1 \qquad &
              \mbox{(elliptic)}\\
             r \qquad & \mbox{for} & \qquad \epsilon = 0 \qquad &
             \mbox{(flat)}\\
           \sinh r \qquad &\mbox{for} & \qquad \epsilon = -1 \qquad &
            \mbox{(hyperbolic)}\ea \right. \  .
\ea                                                     \label{robw}
\ee
The spatial part has a constant curvature and its geometry
is determined by $\epsilon $: flat
($\epsilon = 0$), elliptic ($\epsilon = +1$) or hyperbolic ($\epsilon = -1$).
The flat and hyperbolic cases correspond to open universes
whereas the elliptic case
corresponds to a closed universe. The whole dynamics of this metric is
contained in the world radius $K(\tau)$ which has to be determined
by the field equations.

First we demonstrate how to get a cosmological solution
for $\epsilon=1$. In this case the spatial part of the space time
is a $S_3$ manifold with $K(\tau)$ as the time dependent radius.
The most general 5d metric respecting the $S_3$ symmetry
is given by a Schwarzschild metric which can be written as
\be
\tilde{ds}^2 = e^{\nu(t)} dx^2 - e^{\lambda(t)} dt^2 + t^2
d\Omega_{3,\epsilon =1}^{2}
\ee
where $x$ is our fifth coordinate which the theory should not depend on
and $t$ corresponds to the time in the 4d theory. In comparison
to the usual Schwarzschild metric our time corresponds to the radius
and $x$ to the time, however, with opposite signs in front of $dx^2$ and
$dt^2$. Because we have no matter in the 5d theory a nontrivial vacuum
solution satisfying the desired $S_{3}$ symmetry is given by a 5d Black Hole
\be
e^{-\lambda} = e^{\nu} = -1 + \frac{2m}{t^2},
\ee
where $m$ is an integration constant. The
properties of this metric are well known. There is a singularity
at $t = 0$ and for $m > 0$ we have a horizon at $t^2 = 2m$.

The generalization of this solution for arbitrary
$\epsilon $ is given by
\footnote{In this case we have to replace $\sin r$ in $d\Omega_{3,1}^2$
by $\frac{\sin (\sqrt{\epsilon} r)}{\sqrt{\epsilon}}$ in
$d\Omega_{3,\epsilon}^2$}
\be                                                \label{solution}
e^{-\lambda} = C \, e^{\nu} = -\epsilon + \frac{2m}{t^2}
\ee
and a horizon appears therefore for $\frac{m}{\epsilon} > 0$
($\epsilon \not=0$). Let us now
perform the reduction to the 4d theory. In terms of (3) it is easy
to obtain the dilaton field and the 4d metric
\be        \label{strfr}
\ba{rcl}
\nu&=&(\alpha + \beta) \phi\\
ds^2&=&e^{-\beta \phi} \left( - e^{\lll} dt^2 + t^2 d\Omega_3^2 \right)\\
  &=&- \left(e^{\lll}\right)^{\frac{\alpha+2\beta}{\aaa+\bbb}} \, dt^2 +
\left(e^{\lll}\right)^{\frac{\bbb}{\aaa+\bbb}} \, t^2 \, d\Omega_{3,\eee}^2 ~.
\ea
\ee
Because the exponents of $e^{\lll}$ in (\ref{strfr}) are in general 
not integers (cf.\ (\ref{alf})) it is
impossible to perform analytically the integration
$e^{\frac{\alpha+2\beta}{\aaa+\bbb}\lll} \, dt^2 = d\tau^2$. At the end
we will discuss some special cases and present some numerical results.
Furthermore, in order to have a real metric in (\ref{strfr}) we have the
restriction that (\ref{solution}) has to remain
positive, i.e.\ $\frac{2m}{t^2} > \eee$.

Before discussing the solution (\ref{strfr}) in detail we
consider a possible generalization of the 5d theory. The simplest
extension is given by adding a cosmological constant, i.e.
\be                                                 \label{fivedesit}
\tilde{S} \rightarrow \tilde{S} = \int d^5x \sqrt{|\tilde{G}|}
     \left( \tilde{R} - \LL \right) ~.
\ee
Again we look for a ``static'' BH solution and find for arbitrary
$\eee$
\be                  \label{loesung}
e^{-\lambda} = C \,e^{\nu} = -\eee + \frac{2m}{t^2} + \frac{\LL}{12} t^2 ~.
\ee
For $\eee = 1$ this solution corresponds to the known 5d
Schwarzschild - de Sitter metric \cite{bere} (after interpreting $x$ as time
and $t$ as radius). The constant $C$ can be eliminated by a constant rescaling
of $x$ or equivalently by a constant shift
in the dilaton (cf. (\ref{strfr})), i.e. the constant part of the
dilaton ($\phi \sim \phi_0 + \phi(t)$) is fixed by the $x$ scale.
Another useful parameterization is given by
\be
e^{-\lll} = \frac{\LL}{12} \frac{(t^2 - t_+) (t^2 - t_-)}{t^2}
  \qquad \mbox{with} \qquad t_{\pm} = \frac{6 \eee}{\LL} (1 \pm \sqrt{1 -
  \frac{2}{3}  \frac{m \LL}{\eee^2}}) ~.
\ee
where $t_{\pm}$ are the two horizons (BH and de Sitter) of the
Schwarzschild - de Sitter metric. Both horizons
coincide at the critical limit $3 \eee^2 = 2m \LL$.
In order to get a real 4d metric we have here (as for $\LL = 0$)
the restriction that $e^{\lll} > 0$.
If we reduce the 5d action (\ref{fivedesit}) in terms of (\ref{fuenf})
to the 4d theory
we observe that the cosmological constant produces a dilaton potential in
four dimensions
\be            \label{aevektif}
\tilde{S} = \frac{1}{2} \int d^5x \sqrt{|\tilde{G}|}
  \left( \tilde{R} - \LL \right) \quad \rightarrow \quad S = \frac{1}{2}
  \int d^4 x \sqrt{|G|}\,e^{-2\phi} \left( R - 4\omega (\pa \phi)^2 -
  \LL e^{\bbb \phi}  \right) ~.
\ee
The corresponding 4d metric and dilaton are again given by (\ref{strfr}).
It is easy\footnote{It is especially easy  if one used suitable
computer programs like MATHEMATICA/MATHTENSOR.} to check that these
background fields fulfill the corresponding $\bar{\bbb}$ equations
($\omega = -1$)
\begin{equation}
 \begin{array}{l}
 R_{\mu\nu} + 2 D_{\mu} \pa_{\nu}\phi
 + \LL\, \frac{\bbb}{4}\, e^{\bbb \phi} G_{\mu\nu} = 0 \\
- D^2 \phi + 2 (\pa \phi)^2 - \LL\, \frac{\bbb + 2}{4}\, e^{\bbb \phi} = 0
\ea
\ee
with $\bbb = -2 ( 1 \pm \sqrt{\frac{1}{3}})$. The structure of the dilaton
potential is similar to the higher genus contributions. The main difference,
however, is that in our case the power of $e^{\phi}$ is real whereas
in the higher genus contribution it is an integer \cite{atse}.

\vsf

To get contact with standard cosmology let us now consider the
solution (\ref{loesung})
in the Einstein frame which is defined by the Weyl transformation
\be
G_{\mu\nu}^{(E)} = e^{-2\phi} G_{\mu\nu}
\ee
and for the effective action (\ref{aevektif}) we find
\be
\begin{array}{c}
S=\int d^4 x \sqrt{ |G^{(E)}| } \left[ R^{(E)} + 2 \omega (\pa\phi)^2 -
  \LL \, e^{(\bbb+2) \phi}    \right] ~.
\end{array}
\ee
In this parameterization we have the standard Einstein-Hilbert term
as gravitational part and the matter part is given by the dilaton terms.
The corresponding energy--momentum tensor is
\begin{equation}
T_{\mu\nu}^{matter} = 2 \left( \partial_{\mu} \phi \partial_{\nu} \phi
 - \frac{1}{2} (\partial \phi )^2 G_{\mu\nu}^{(E)} \right) -
 \frac{1}{2} \LL \, e^{(\bbb +2) \phi} \, G_{\mu\nu}^{(E)} ~.
\end{equation}
In this frame the 4d metric is given by
\be
ds_{E}^2 = - e^{\lll /2} dt^2 + t^2 e^{-\lll /2} d\Omega_{3,\eee}^2 ~.
\ee
In contrast to the string frame (\ref{strfr}) the metric does not contain
the root expressions (\ref{alf}).

\vsf

In order to obtain statements about the evolution of the (4d) universe
we have to bring the solution into the form (\ref{robw}),
where the world radius
$K(t(\tau))$ is given by
\be
\ba{ll}
K^2 = t^2\, (e^{\lll})^{\frac{\bbb}{\aaa + \bbb}}  \qquad  &\qquad
 \mbox{(in the string frame)} ~,\\
K_{(E)}^2 = t^2\, e^{-\lll /2}  \qquad & \qquad
 \mbox{(in the Einstein frame)} ~,\\
e^{-\lll} = - \eee +\frac{2m}{t^2} + \frac{\LL}{12} t^2 &
\ea
\ee
and the function $t(\tau)$ is a solution of the differential equation
\be
\ba{ll}
\dot{t}^2(\tau) = (e^{-\lll})^{\frac{\aaa + 2\bbb}{\aaa + \bbb}} \qquad
& \qquad \mbox{(in the string frame)},\\
\dot{t}^2(\tau) = e^{-\lll /2} \qquad & \qquad \mbox{(in the Einstein frame)}.
\ea                                        \label{dgln}
\ee
If there are no horizons (in the 5d theory) it is useful to fix
the integration constant of (\ref{dgln}) via $t(0) = 0 $,
i.e.\ the singularity (big bang) appears at $\tau =0$. In this case
$t$ runs from zero to infinity. The cases where there
are horizons (defined by $e^{-\lll}=0$) we have to restrict ourselves
to regions with $e^{-\lll}>0$ and the horizons correspond either to the
beginning or to the end of the universe.
Extremal values of $K$ possibly occur at $\frac{dK}{d\tau}=0$.
Provided that $\dot t \not= 0$ this is equivalent to
\begin{equation}       \label{extrema}
-\epsilon + (1-2\delta )\frac{2m}{t^2}+\frac{\Lambda}{12}(1+2\delta)t^2=
0,
\end{equation}
where $\delta = - \frac{\beta}{2(\alpha +\beta )} = \frac{1}{4}(\sqrt{3}
\pm 1)$ in the string frame, and $\delta =\frac{1}{4}$ in the Einstein frame.
Since we were not able to solve (\ref{dgln}) in general we are going to
consider now the asymptotic behavior of the world radius.

${\bf 1.\quad t^2 << \min (2m,\frac{1}{|\Lambda |}),\, m>0}$:\
We emphasize that this limit makes sense only if
$\alpha^{\prime} << t^2 $ is still valid since we
have neglected higher order corrections in $\alpha^{\prime}$.
The zeroth order asymptotic behavior of $K$ is given by
\begin{equation}       \label{mueller}
\ba{ccc}
K^2(\tau ) \sim \tau ^{\mp \frac{2}{\sqrt{3}}} &\quad , \quad &
K_{E}^2(\tau ) \sim \tau ^{\frac{2}{3}}\\
e^{2\phi}\sim\tau^{-(1\pm \sqrt{3})}     &\quad , \quad &
e^{2\phi_{E}}\sim\tau^{\mp \frac{2\sqrt{3}}{3}} ~.
\ea
\end{equation}
Although the dilaton is not transformed if we go to the Einstein frame
we have to perform different time redefinitions in both frames.
For $\epsilon =\Lambda =0$ (\ref{mueller}) is an exact expression.
This is just the solution which has already been given
by Mueller in \cite{muell}.
Another interesting asymptotic region is

${\bf 2.\quad t \rightarrow \infty \, (\Lambda \ge 0):}$\
In this limit we find the following behavior
\be                                    \label{unend}
\ba{ccc}
K^2(\tau ) \sim \left\{
\ba{cc}
\tau^{\pm 2\sqrt{3}} & \LL \not= 0 \\
\tau^2 & \LL = 0 \ea \right.
 & \quad , \quad &
K_{E}^2 \sim \left\{
\ba{cc}
\tau^{6} & \LL \not= 0 \\
\tau^2 & \LL = 0 \ea \right. \\
e^{2\phi}\sim \left\{
\ba{cc} \tau^{3\mp\sqrt{3}} & \LL \not= 0\\
1 & \LL =0 \ea \right.
   & \quad , \quad &
e^{2\phi_{E}}\sim \left\{
\ba{cc} \tau^{\pm 2\sqrt{3}}& \LL \not= 0 \\
1 & \LL = 0 \ea \right.  ~.
\ea
\ee
For vanishing potential ($\LL =0$) we have to restrict ourselves
on $\epsilon =-1$ ($e^{\lambda} > 0$). In this case we get the 
remarkable consequence that in the asymptotic limit our solution
is in both frames a flat space time ($K=\tau$) with
constant dilaton.

If there are horizons in the five dimensional theory we have to consider
also the asymptotics near those horizons.

\footnotesize
\begin{table}[t] \centering
\begin{tabular}{|r|r|r|r|r|r|r|r|}  \hline
$\epsilon$ & $m$ & $\Lambda$ & ``horizon'' & extrema &
$t\rightarrow 0$ & $t\rightarrow$ ``horizon'' & $t\rightarrow
\infty $                                                      \\[-0.5ex]
 & & & &possible for &$K \sim$ &$K\sim$&$K\sim   $        \\ \hline \hline
$0$ & $>0$ & $0$ & no & --
   &$ \tau ^{\mp\frac{\sqrt{3}}{3}}$
   & -- &
   $\tau ^{\mp\frac{\sqrt{3}}{3}}$                               \\ \hline
$0$ & $0$ & $ >0$ & no & -- &$\tau^{\pm\sqrt{3}}$&--
   &$\tau^{\pm\sqrt{3}}$                                          \\ \hline
$0$&$>0$&$>0$&no&upper
sign&$\tau^{\mp\frac{\sqrt{3}}{3}}$&--&$\tau^{\pm
\sqrt{3}}$                                      \\ \hline
$0$&$<0$&$>0$&$t=\left( -\frac{24m}{\Lambda} \right)^{\frac{1}{4}}$&
   --&--&$\tau^{\pm\frac{\sqrt{3}}{3}}$&
   $\tau^{\pm\sqrt{3}}$                                \\ \hline
$0$&$>0$&$<0$&$t=\left( -\frac{24m}{\Lambda}\right)^{\frac{1}{4}}$&
   lower sign&$\tau^{\mp\frac{\sqrt{3}}{3}}$&
   $(\tau_{0}-\tau )^{\pm\frac{\sqrt{3}}{3}}$&--                    \\ \hline
$-1$ & $>0$ &$0$ & no & upper sign
   &$ \tau ^{\mp\frac{\sqrt{3}}{3}}$ & -- & $\tau$        \\ \hline
$-1$&$0$&$>0$&no&--&$\tau$&--&$\tau^{\pm\sqrt{3}}$    \\ \hline
$-1$ &$ >0$ &$>0$&no&upper sign&$\tau^{\mp\frac{\sqrt{3}}{3}}$&
   --&$\tau^{\pm\sqrt{3}}$                                \\ \hline
$-1$ &$>0$ &$ <0$&$t^2=t_{+}$&both sign&
   $\tau^{\mp\frac{\sqrt{3}}{3}}$&$(\tau_{0}-\tau )^{\pm\frac{\sqrt{3}}{3}}$&
   --                                                               \\ \hline
$-1$ &$<0 $& $>0$&$t^2=t_{-}$&--&--&
   $\tau^{\pm\frac{\sqrt{3}}{3}}$&$\tau^{\pm\sqrt{3}}$ \\ \hline
$^* -1$ &$<0$&$<0$&$t^2=t_{-}$&upper sign&--&$\tau^{\pm\frac{\sqrt{3}}{3}}$
  &--                               \\[0.2ex]
  & & &$t^2=t_{+}$& & &$(\tau_{0}-\tau )^{\pm\frac{\sqrt{3}}{3}}$& \\ \hline
$-1$ &$<0$&0&$t^2=-2m$&lower sign&--&
  $(\tau_0-\tau )^{\pm\frac{\sqrt{3}}{3}}$ & $\tau^{\pm\sqrt{3}}$ \\ \hline
$-1$&0&$<0$&$t^2=-\frac{12}{\LL}$& upper signs&$\tau$& $(\tau_0-\tau
  )^{\pm\frac{\sqrt{3}}{3}}$ &-- \\ \hline
$1$ &$>0$ &0 &$t^2=2m$ &--&$\tau^{\mp\frac{\sqrt{3}}{3}}$&$(\tau_0-
  \tau)^{\pm \frac{\sqrt{3}}{3}}$&--   \\ \hline
$1$&$0$ &$>0$ &$t^2=\frac{12}{\LL}$
   &--&--&$\tau^{\pm\frac{\sqrt{3}}{3}}$&$\tau^{\pm\sqrt{3}}$ \\ \hline
$^*1$ &$>0$ &$>0$&$t^2=t_-$&lower sign&$\tau^{\mp\frac{\sqrt{3}}{3}}$&
   $(\tau_0 - \tau)^{\pm\frac{\sqrt{3}}{3}}$&--\\  \cline{6-8}
   & & &$t^2=t_+$& &-- &$\tau^{\pm\frac{\sqrt{3}}{3}}$&
  $\tau^{\pm\sqrt{3}}$\\ \hline
$1$&$>0$&$<0$&$t^2=t_-$&lower
   sign&$\tau^{\mp\frac{\sqrt{3}}{3}}$&$(\tau_0-\tau
   )^{\pm\frac{\sqrt{3}}{3}}$&--                         \\ \hline
$1$ &$<0$ &$>0$&$t^2=t_+$&upper
  sign&--&$\tau^{\pm\frac{\sqrt{3}}{3}}$&$\tau^{\pm\sqrt{3}}$
  \\ \hline
\end{tabular}
\caption{Collection of the solutions in the string frame. ($^{*}
m\Lambda <\frac{3}{2}$)}  \label{saite}
\end{table}

\normalsize

${\bf 3.\quad t\rightarrow \mbox{\bf
``horizon''}}(\frac{2m}{\epsilon}>0,\, \LL =0)$\footnote{Here the
restriction on
$\LL =0$ is done in order to simplify the discussion, a
consideration of $\LL \not= 0$ is also possible (cf. table
\ref{saite}).}:\ In the previous
cases we admitted $t$ to run from zero to infinity. However, if there are
horizons in the five dimensional theory we have to care
that our solution is real ($e^{\lambda}>0$). First we will consider the case
$\epsilon = -1$. The metric
is real if $t\in(\sqrt{\frac{2m}{\epsilon}},\infty )$. We fix the integration
constant of (\ref{dgln}) via
\begin{equation}                            \label{anfbed}
t(0)=\sqrt{\frac{2m}{\epsilon}},
\end{equation}
which ensures that $\tau $ runs from zero to infinity.
With (\ref{anfbed}) we obtain the behavior
\be
\ba{lcl}
K^2(\tau ) \sim \tau ^{\pm \frac{2}{\sqrt{3}}} & \quad , \quad &
 K_{E}^2(\tau ) \sim \tau ^{\frac{2}{3}}\\
e^{2\phi}\sim\tau^{-(1 \mp\sqrt{3})}  & \quad , \quad &
e^{2 \phi_{E}}\sim  \tau^{\pm\frac{2\sqrt{3}}{3}}  ~.
\ea
\ee
In the other case ($\epsilon =+1$) the metric is real
if $t \in (0, \sqrt{\frac{2 m}{\epsilon}})$, i.e.\ the
lifetime of the universe is finite. In that case the behavior
at zero is given by (\ref{mueller}) and near the horizon by
\be
\ba{lcl}
K^2(\tau ) \sim (\tau_{0} -\tau )^{\pm \frac{2}{\sqrt{3}}} &\quad ,\quad &
K_{E}^2(\tau ) \sim (\tau_{0} -\tau )^{\frac{2}{3}}\\
e^{2\phi}\sim(\tau_0-\tau )^{-(1\mp\sqrt{3})} &\quad , \quad &
e^{2\phi_{E}}\sim (\tau_0-\tau )^{\pm\frac{2}{\sqrt{3}}}  ,
\ea
\ee
where $\tau \rightarrow \tau_{0}$ at the ``horizon''.

We will not discuss every possible parameter constellation in detail but
collect the solutions in table \ref{saite} for the string frame. With
the given data the same can be done for the Einstein frame very
easily.

Finally we give the result of numerical calculations done in some
special cases. \mbox{Figure 1} is an example for a closed universe 
($\eee =1$) with finite lifetime and vanishing dilaton potential.
The expansion starts at $\tau =0$, reaches the maximum at $t^{2}(\tau ) =m$,
(cf.\ (22)) and shrinks to zero size at the ``horizon'' $t^2(\tau )=2m$.
The corresponding ``lifetime'' of the universe is according to (\ref{dgln})
given by
\be
\tau = \int_0^{\sqrt{2m}} \left(\frac{t^2}{2m -
  t^2}\right)^{\frac{q}{2}} dt =
    \sqrt{\frac{2m}{\pi}}\, \Gamma(\frac{q+1}{2}) \, \Gamma(1-\frac{q}{2})
\ee
where: $q=\frac{1}{2}(1 \mp \sqrt{3})$ in the string frame and
$q = \frac{1}{2}$ in the Einstein frame. \mbox{Figure 2} shows an open
(hyperbolic) universe expanding forever with decreasing velocity.
The asymptotic limit is a flat space time with a constant dilaton.
In figure 3 the ``big bang'' appears as an implosion at $t=0$, the universe
shrinks until it reaches it's minimum at $t^4(\tau )=\frac{24m}{\LL}
\frac{2\delta - 1}{2\delta + 1}$ (cf.\ (22)) and then expands with increasing
velocity. This is an example for an open universe which is spatially flat
($\eee = 0$). The asymptotic behavior is given by (\ref{unend})
and we see that the dilaton potential ($\LL \neq 0$) accelerates
the expansion in both frames.

\begin{figure}[t] \vspace*{-25mm}
\begin{minipage}[t]{7.3cm}
\mbox{\epsfig{file=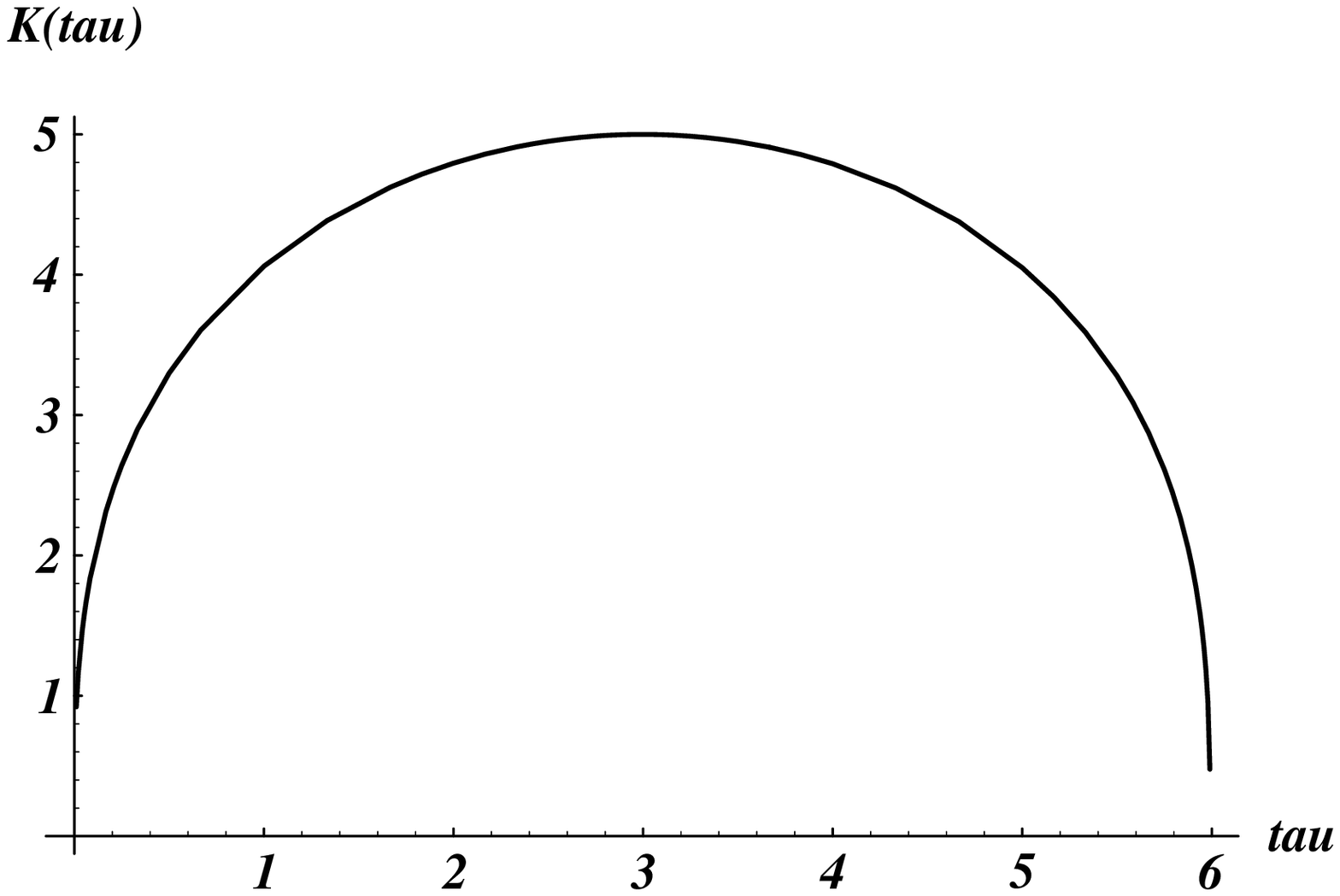,height=10cm,width=7cm}} \vspace{-25mm} \par
\caption{$K(\tau )$ in the Einstein frame for $\epsilon=1$, $m=50, \LL =0$}
 \end{minipage}
 \hfill
 \begin{minipage}[t]{7.3cm}
 \mbox{\epsfig{file=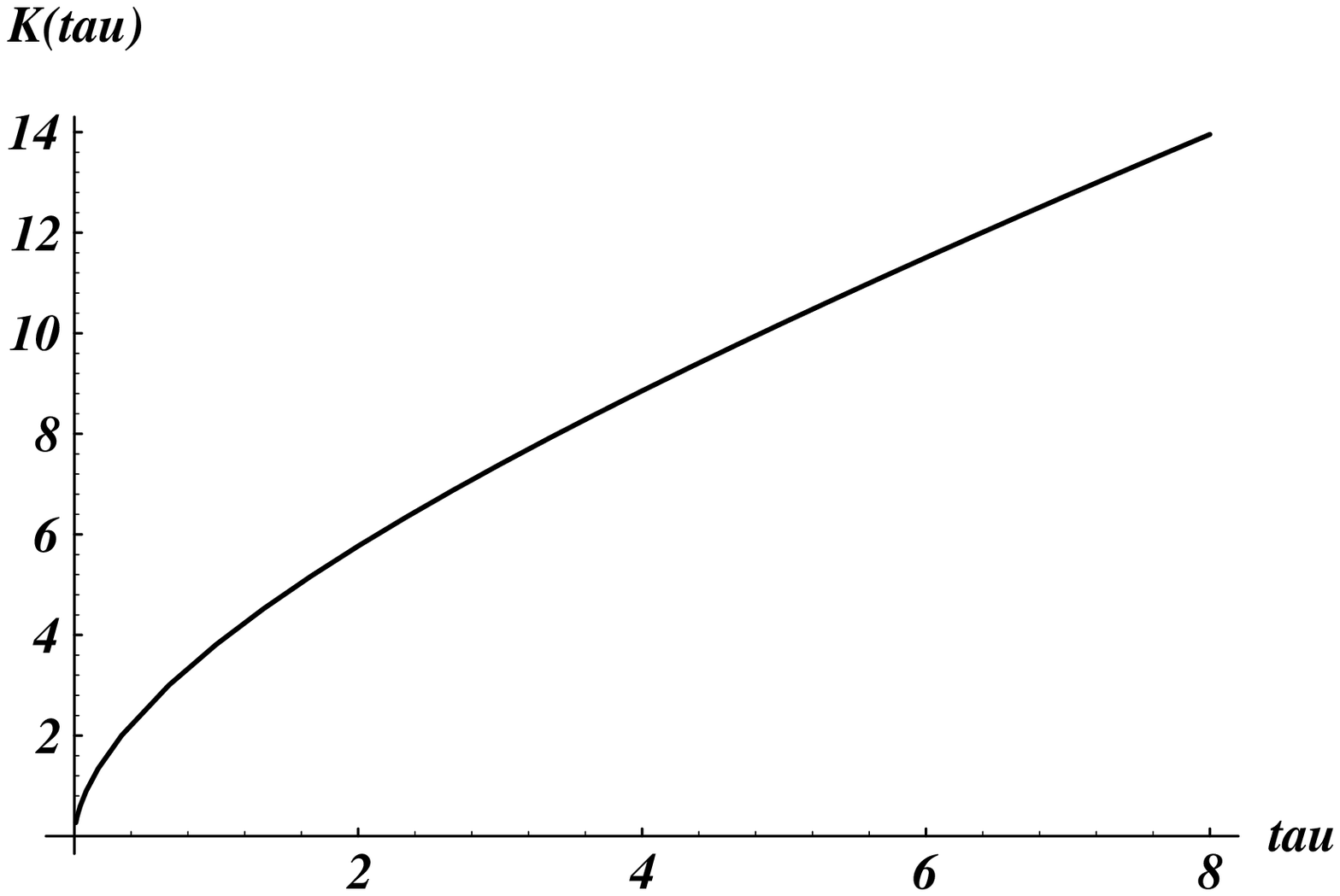,height=10cm,width=7cm}} \vspace{-25mm} \par
\caption{
$K(\tau )$ in the string frame for $\epsilon=-1$, $m=50,\LL =0$}
 \end{minipage} \vspace{-3cm}
\label{bild1}
\hfill \begin{minipage}[t]{7.3cm} \hfill
 \mbox{\epsfig{file=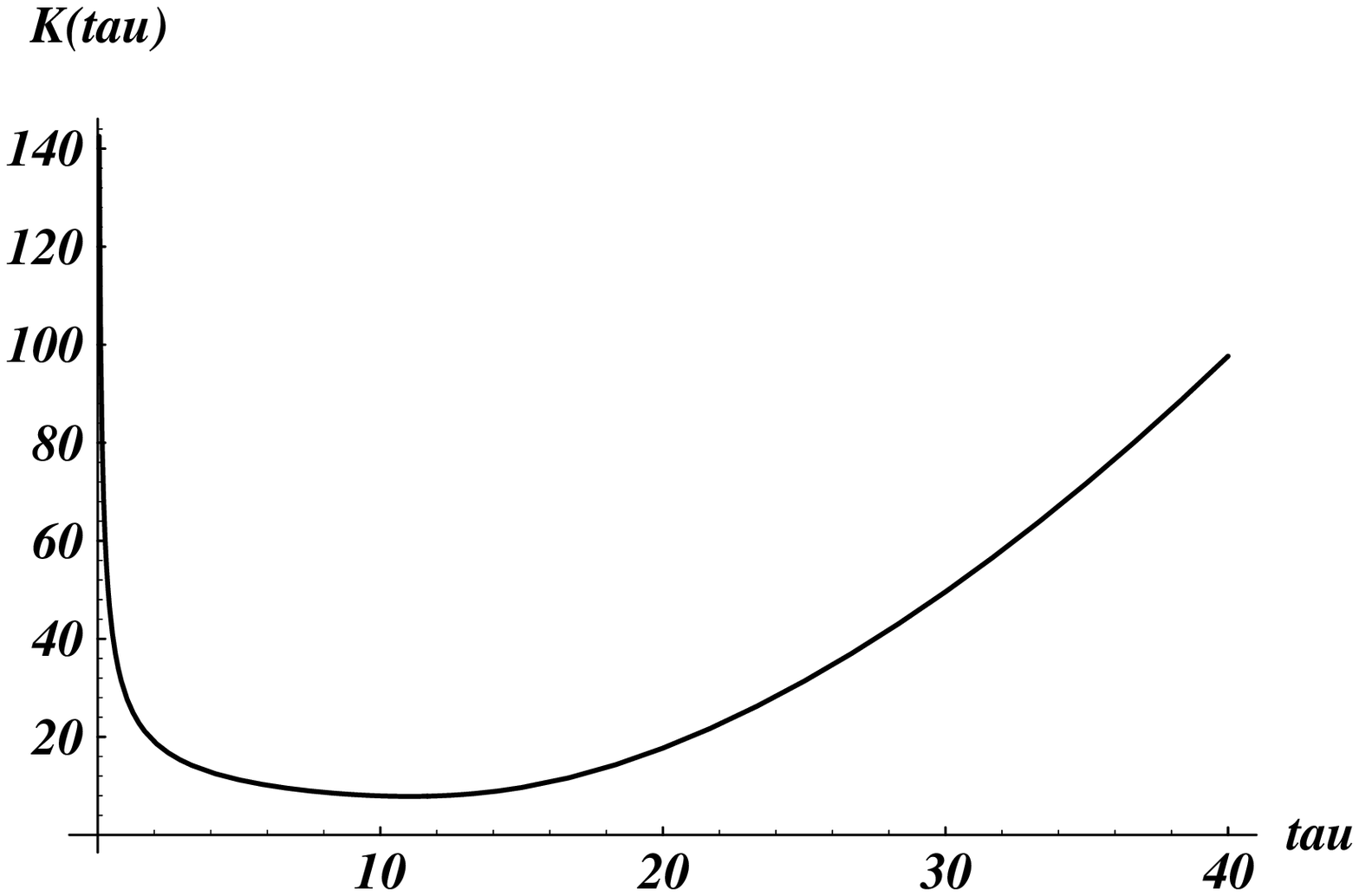,height=10cm,width=7cm}} \vspace{-25mm}
\caption{
$K(\tau )$ in the string frame for $\epsilon=0$,
 $ m=50,\LL =1/m$}   \hfill
\end{minipage}
\label{bild3} \vspace*{-10mm}
\end{figure}

\vsf

To summarize, in the present paper we have obtained various cosmological
solutions of the low energy effective action of string theory.
Although the method presented is very simple (reduction of a 5d Black
Hole solution) our solution has as far as we know not been obtained before.
The reason is that  we did not use the standard parameterization
of the Robertson-Walker metric (\ref{robw}) and it seems to be impossible
to solve (\ref{dgln}) analytically. However,
in most cases it is possible to get some impression about the features
of our solutions. We have investigated the asymptotic behavior
of the world radius near
zero, near the horizons of the corresponding 5d Black Hole, and in the
infinite future. For vanishing dilaton potential and $\epsilon = -1$ we
obtained a flat universe in the large time limit.
In the case where we were able to get an analytic expression ($\LL =
\epsilon =0$) in the Robertson-Walker parameterization (\ref{robw})
our result coincides with Mueller's solution
\cite{muell}. Finally we should mention that the cosmological scenarios
implemented by our solutions have not to be taken too seriously since
one can not expect to get quantitative exact results in such a simple
model (pure dilaton, graviton system). Therefore it would be very
useful to take into account additional matter. Unfortunately the
incorporation of matter is not straightforward because this simple
reduction  procedure works only for the pure dilaton graviton system.
On the other hand it should be possible by duality and/or O(d,d)
transformations to construct new solutions in the 5d theory and perhaps
it is possible to interpret the corresponding 4d fields in the
context of string theory.

\vspace{8mm}

\noindent
{\large\bf Acknowledgments} \vspace{3mm} \newline \noindent
We would like to thank H.\ Dorn, D.\ L\"ust and G.\ Weigt
for helpful discussions.

\renewcommand{\arraystretch}{1.0}

\end{document}